\documentclass[pra, twocolumn, a4paper, nofootinbib, showpacs, showkeys]{revtex4}

\usepackage{graphicx}
\usepackage[bookmarks=true]{hyperref}
\usepackage{amsmath}
\usepackage{amssymb}
\usepackage{lipsum}
\usepackage{pgfplots}
\usepackage{adjustbox}
\usepackage[utf8]{inputenc}

\newcommand{\ket}[1]{\vert{#1}\rangle}
\newcommand{\bra}[1]{\langle{#1}\vert}
\newcommand{\Tr}{\textnormal{Tr}}

\newcommand{\abs}[1]{\left| #1\right|}
\newcommand{\norm}[1]{\left\| #1\right\|}

\newcommand{\Cc}{{\mathbb{C}}}

\newcommand{\Nn}{{\mathbb{N}}}

\def\tensor{{\otimes}}

\renewcommand{\emph}{\textit}
\begin{document}

\title{Pretty good state transfer of entangled states through quantum spin chains}

\date{September 29, 2014}

\author{R\'uben Sousa}
\affiliation{Physics of Information Group, Instituto de Telecomunica\c{c}\~oes, P-1049-001 Lisbon, Portugal}
\affiliation{CEMAPRE, ISEG, Universidade de Lisboa, P-1200-781 Lisbon, Portugal}

\author{Yasser Omar}
\affiliation{Physics of Information Group, Instituto de Telecomunica\c{c}\~oes, P-1049-001 Lisbon, Portugal}
\affiliation{CEMAPRE, ISEG, Universidade de Lisboa, P-1200-781 Lisbon, Portugal}

\begin{abstract}
The XX model with uniform couplings represents the most natural choice for quantum state transfer through spin chains. Given that it has long been established that single-qubit states cannot be transferred with perfect fidelity in this model, the notion of \textit{pretty good state transfer} has been recently introduced as a relaxation of the constraints on fidelity. In this paper, we study the transfer of multi-qubit entangled and unentangled states through unmodulated spin chains, and we prove that it is possible to have pretty good state transfer of any multi-particle state. This significantly generalizes the previous results on single-qubit state transfer, and opens way to using uniformly coupled spin chains as short-distance quantum channels for the transfer of arbitrary states of any dimension. Our results could be tested with current technology.
\end{abstract}



\maketitle

\section{Introduction}

The transfer of quantum states from one site to another is a key task in quantum information processing. For example, one would like to have a quantum data bus between quantum registers and/or processors capable of transmitting arbitrary quantum states in a reliable way \cite{bose2}. A straightforward approach could be to apply a sequence of SWAP gates, but that would be too demanding in terms of control, and very prone to errors \cite{swaperror}. This has stimulated the proposal of passive quantum networks as transmission devices which do not require control except during the preparation and readout. In particular, the one-dimensional half-integer spin chain has been extensively studied as a quantum wire for the transfer of qubit states \cite{bose1, spin1/2, state}.

Much attention has been given to the ideal scenario where we have \textit{perfect state transfer} (PST), i.e.\ where the fidelity of transfer is equal to one. Such unit fidelity has been investigated in the context of a wide range of quantum spin networks (see \cite{kayreview} for a review), and it has been demonstrated that PST cannot be achieved within spin-1/2 chains with basic nearest-neighbor couplings. The experimental realization of spin chains with PST would be possible if the system was engineered according to non-natural coupling schemes \cite{couplings1, couplings2, couplings3}, but this would unfortunately be a highly difficult task in practice. Furthermore, it has been argued that the condition of perfect transfer is far too demanding in comparison with the level of fidelity required for the implementation of most quantum information processing tasks \cite{information}. Therefore, it is important to investigate if a relaxation of the constraints on the fidelity can lead to better results.

It would be useful to establish that quantum state transfer can be accomplished in chains with minimal variation of the coupling strengths. This is so because the unmodulated chain is much easier to fabricate as compared to chains where local engineering is required \cite{bose2} and, furthermore, there exist experimental quantum information transfer platforms whose natural dynamics are governed by Hamiltonians with uniform couplings \cite{expuniform, nanomagnets}. Motivated by this, several alternatives have been proposed, such as designing weakly varying coupling configurations which also display the PST property \cite{spinmirror}, obtaining high quality ballistic state transfer or dynamic generation of entanglement in chains where the bulk is uniform and the couplings between the bulk and the boundary qubits are tuned to optimal values \cite{minengineer1, minengineer2, minengineer3, minengineer4}, using an iterative measurement procedure to perform state transfer in unmodulated chains with low probability of failure \cite{iterative1, iterative2}, and improving the quality of the transmission by initializing the channel qubits to a specific state \cite{init}.

In our work, we focus on the notion of \textit{pretty good state transfer} (PGST), which has recently been introduced as a significant alternative to PST \cite{first, apst}. Here, the requirement is that the fidelity of transfer gets arbitrarily close to unity. By applying this concept to the unmodulated XX-type chain, which represents the model of spin chain whose construction requires less control over the individual parts \cite{bose2}, Godsil \textit{et al.\ }proved that there is PGST of single-particle states in such a chain if and only if the length of the chain is $n=p-1$, $n=2p-1$, where $p$ is prime, or $n=2^k-1$ \cite{pgst}. This result means that the quality of the state transfer protocol depends on conditions which are purely number-theoretic, and shows that there is a surprising connection between the dynamics of quantum spin chains and primality.

Since the discovery by Shor of a polynomial time algorithm for prime factorization on a quantum computer \cite{shor}, the application of quantum algorithms to problems in prime number theory has been a topic of special interest. This continues to be an active research direction: for instance, a recent contribution is that of Latorre and Sierra, who propose the creation of a single quantum state made of the superposition of prime numbers to study primality problems \cite{latorre}. In this context, the result of \cite{pgst} is of high significance, since it establishes a nontrivial link between quantum dynamics and primality which is outside the scope of the traditional algorithmic applications. It would therefore be desirable to generalize this result, namely to other protocols with practical relevance for quantum state transfer. 

The work in \cite{pgst} is restricted to single-particle qubit states. However, given the practical challenges inherent to the implementation of quantum systems, the usefulness of the experimental realization of the uniformly coupled XX chain as a quantum channel clearly depends on whether or not many-particle qubit states, and in particular entangled states, can also be transferred with arbitrarily high fidelity through a single chain. This topic has already been addressed for PST protocols \cite{spinmirror, multiperturbation}, and it has been established that, in the XX model, PST of arbitrary single-qubit states is a sufficient condition for PST of multi-qubit states\footnote{Thanks to the Jordan-Wigner transformation, PST of single-particle states suffices for perfect mirror inversion of quantum states with any number of excitations, modulo the generation of phase factors \cite{kayreview}. Such phase factors can be corrected via a properly constructed quantum gate \cite{phasegate}.}. While completing this manuscript, we also became aware of an independent work in chains with Rabi-like dynamics \cite{manyqubit} that further endorses the relevance of the problem of multi-particle state transfer through a single chain. However, to the best of our knowledge, the study of PGST has never encompassed multi-particle state transfer.

In this paper, we extend the definition of PGST to arbitrary multi-qubit states, and we show that regardless of the number of qubits of the input state, the fidelity of the transfer is arbitrarily high if and only if the length of the chain is $n=p-1$, $2p-1$, or $2^k-1$. From a strictly physical point of view, this confirms that the uniformly coupled quantum spin chain is a versatile channel for the construction of quantum communication systems. Additionally, our work widens the number of protocols for which PGST is characterized as a function of primality conditions. This highlights that the link between quantum dynamics and primality goes beyond the established applications in the field of quantum algorithms. 

\section{State transfer of $m$-qubit states}

We consider a model of $n$ qubits interacting within a one-dimensional spin-1/2 system described by an XX-type Hamiltonian with isotropic couplings,
\begin{equation} \label{eq:Hfirst}
H={1 \over 2} \sum_{j=1}^{n-1} J \left( \sigma_{j}^{x}\sigma_{j+1}^{x} + \sigma_{j}^{y}\sigma_{j+1}^{y} \right),
\end{equation}
where $\sigma_{j}^{x}, \sigma_{j}^{y}$ and $\sigma_{j}^{z}$ are the Pauli spin operators at position $j$. Without loss of generality, we will consider $J=1$.

By employing the Jordan-Wigner transformation \cite{jordwig}, this system can be mapped to a local fermionic Hamiltonian,
\begin{equation} \label{eq:Hferm}
H={1 \over 2} \sum_{j=1}^{n-1} \left( c_{j}^{\dagger}c_{j+1} + c_{j+1}^{\dagger}c_{j} \right),
\end{equation}
with
\begin{equation}
c_{j} = \left( \prod_{l<j} \sigma_{l}^{z} \right) {\sigma_{j}^{x} + i\sigma_{j}^{y} \over 2},
\quad
c_{j}^{\dagger} = \left( \prod_{l<j} \sigma_{l}^{z} \right) {\sigma_{j}^{x} - i\sigma_{j}^{y} \over 2}.
\end{equation}
The total $z$-spin operator, given by $S_{tot}^{z} = \sum_{j=1}^{n} \sigma_{j}^{z}$, commutes with the Hamiltonian of the system, $\left[ H,S_{tot}^{z} \right] = 0$; therefore, the Hilbert space of the register can be diagonalized and decomposed into invariant subspaces consisting of the distinct eigenstates of $S_{tot}^{z}$. Each of these subspaces can be characterized by the number of spins on the excited state, i.e.\ the number of qubits having bit value 1. Since the Hamiltonian \eqref{eq:Hferm} describes a system of non-interacting spinless fermions, initial quantum states that are in each of the individual subspaces of $H$ will remain there under time evolution.

The state transfer scheme we will consider is a natural generalization of the usual single-qubit scheme: the state sender $S$, located on one end of the chain, wishes to transmit an $m$-qubit composite state $\ket{\psi_{in}}$ (which will generally consist of an entangled state) to the receiver $R$, located on the other end (see Fig.\ 1). A generic state of the system is therefore of the form 
\begin{equation} \begin{aligned}
\ket{\Psi} & = \ket{\psi_S} \: \tensor \: \ket{\psi_C} \: \tensor \: \ket{\psi_R}  \\ & = \ket{\psi_1 \psi_2 \ldots \psi_m}  \tensor \ket{\psi_{m+1} \psi_{m+2} \ldots \psi_{n-m} } \\ & 	\qquad \qquad \qquad \quad \tensor \ket{\psi_{n-m+1} \psi_{n-m+2} \ldots \psi_{n} },
\end{aligned} \end{equation}
where $\ket{\psi_S}$ and $\ket{\psi_R}$ belong to the sender and receiver's subspaces respectively, and $\ket{\psi_C}$ belongs to the subspace which corresponds to the rest of the quantum channel. After initialization of the $n$ spins to the eigenstate
\begin{equation}
\ket{\underline{0}} = \ket{0_1 0_2 0_3\ldots 0_{n-1} 0_n} ,
\end{equation}
the sender places the state $\ket{\psi_{in}}$ at the beginning of the chain, and such state shall be recovered at the end of the chain by the receiver after a given time $\tau$.
\begin{figure}[h] \label{figure1}
     \begin{center}
     \resizebox{8.75cm}{!}{\includegraphics{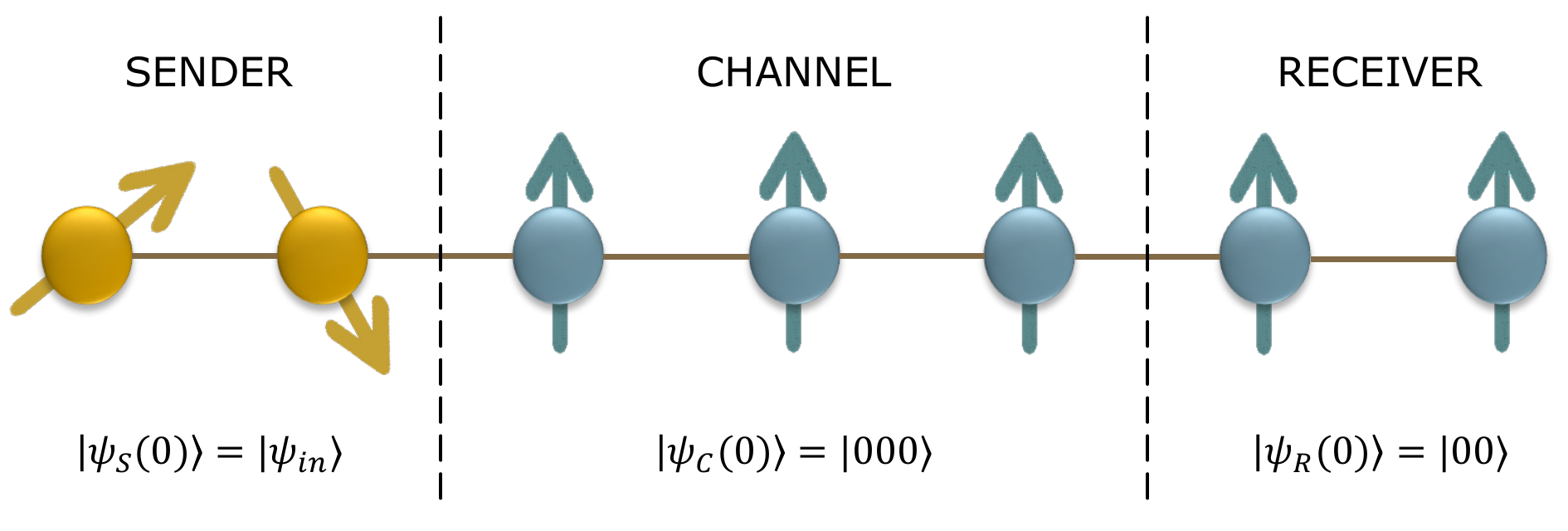}}
     \caption{Diagram of our spin chain state transfer protocol, for the case of two qubit states ($m=2$) and 7 sites ($n=7$).}
     \end{center}
\end{figure}

Since he is located on the antipodal site, we assume that the receiver reads the state of his subsystem in the opposite order. If we let $\ket{\widetilde{\Psi}} = \ket{ \psi_n \ldots \psi_2 \psi_1 }$ be the mirrored state of the whole system, the output state is thus given by
\begin{equation}
\begin{aligned}
\rho_{out}(\tau) & = \Tr_{C,S} \left\{ \ket{\widetilde{\Psi}(\tau)} \bra{\widetilde{\Psi}(\tau)} \right\} \\
& = \Tr_{C,S} \left\{ \ket{\psi_n \ldots \psi_2 \psi_1(\tau)} \bra{\psi_n \ldots \psi_2 \psi_1(\tau)} \right\},
\end{aligned}
\end{equation}
where the partial trace is over the qubits $\ket{\psi_{n-m} \ldots \psi_2 \psi_1}$. (As it will be seen later, this reversal enables us to exploit the symmetry properties of the system.) The figure of merit for the quality of the state transfer through this protocol at time $\tau$ is thus given by the quantum fidelity $\bra{\psi_{in}} \rho_{out}(\tau) \ket{\psi_{in}}$.

%=================================================================

\subsection*{Single-excitation manifold} We will start by restricting our problem to the single-excitation subspace, since this represents the simplest scenario for transfer of entangled states. This subspace is described by the class of states $\ket{j} = \ket{0_1 0_2 \ldots 0_{j-1} 1_j 0_{j+1} \ldots 0_n}$, where only the $j$-th qubit is in the excited $\ket{1}$ state ($j=1,\ldots n$). We are thus assuming that the state of the whole system immediately after the sender places its state on the chain is of the form
\begin{equation} \begin{aligned}
\ket{\Psi (0)} & = \ket{\psi_{in}} \ket{0_{m+1} 0_{m+2} \ldots 0_{n}} \\ & = \alpha \ket{\underline{0}} + \sum_{j=1}^{m} \beta_j \ket{j},
\end{aligned} \end{equation}
where $\abs{ \alpha }^2 + \sum_{j=1}^{m} \abs{ \beta_j }^2 = 1$.

The two definitions of PGST we will adopt in this paper are straightforward generalizations of the notion introduced by Godsil in \cite{first}. We start by stating the first, which applies to the single-excitation scenario: we say that there is PGST of the state $\ket{\psi_{in}}$ if for any $\epsilon > 0$, there is a time $t>0$ such that
\begin{equation} \label{eq:PGSTdef1}
\abs{ \bra{n+1-j} U(t) \ket{j} } > 1-\epsilon, \qquad j=1,\ldots, m,
\end{equation}
where $U(t)=e^{iHt}$ is the time evolution operator. The effects of the absolute values taken above can be easily corrected through the application of appropriate phase gates \cite{phasegate}; therefore, condition \eqref{eq:PGSTdef1} implies that our figure of merit $\bra{\psi_{in}} \rho_{out}(t) \ket{\psi_{in}}$ can be made arbitrarily close to $1$.

Under the single-excitation framework, the state vectors $\ket{j}$ are equivalent to the $n$ vectors of the usual basis of the space $\Cc^n$, and the unitary operator $U(t) = e^{-iHt}$ is equivalent to a $n \times n$ continuous-time quantum walk on the uniformly coupled path graph. This enables us to undertake a graph-theoretic approach in the study of PGST, as in \cite{pgst, universal}. The following theorem is the main result of \cite{pgst}:

\emph{Theorem 1.} Suppose the Hamiltonian of the chain is given by \eqref{eq:Hfirst}. Then, given an $\epsilon > 0$, there is a time $t>0$ such that
\begin{equation} \label{eq:PGST1qubit}
\abs{ \bra{n} U(t) \ket{1} } > 1-\epsilon
\end{equation}
if and only if $n=p-1$, $2p-1$ or $2^k-1$, where $p$ is a prime and $k \in \Nn$.

We will now invoke a particular case of a theorem which was proved by Cameron et al.\ (see Theorem 3 in \cite{universal}).

\emph{Theorem 2.} Suppose the Hamiltonian of the chain is given by \eqref{eq:Hfirst}. Let $F$ denote the $n\times n$ permutation matrix such that $F\ket{j} = \ket{n+1-j}$, for all $j=1, \ldots, n$. Then, condition \eqref{eq:PGST1qubit} is true if and only if for any given $\epsilon >0$ there is a time $t>0$ such that \begin{equation} \label{eq:norm}
\norm{U(t) - \gamma F} < \epsilon
\end{equation} for some $\gamma$ lying on the complex unit circle, where $\norm{\cdot}$ denotes the usual matrix norm.

\emph{Corollary 3.} If the multi-qubit input state $\ket{\psi_{in}}$ is restricted to the single-excitation manifold, then there is PGST of $\ket{\psi_{in}}$ if and only if $n=p-1$, $2p-1$ or $2^k-1$, where $p$ is a prime and $k \in \Nn$.

\emph{Proof.} If $n=p-1$, $2p-1$ or $2^k-1$, then by theorem 2, for each $\epsilon >0$ there is a time $t>0$ such that $\norm{U(t) - \gamma F} < \epsilon$. Therefore,
\begin{equation}
\begin{aligned}
\epsilon & > \max_{\norm{ \ket{\psi} }=1} \norm{ \left(U(t) - \gamma F\right) \ket{\psi} } \\
& \geq  \max_{\ket{k}} \norm{ \left(U(t) - \gamma F\right) \ket{k} } \\
& = \max_{\ket{k}} \norm{ U(t)\ket{k} - \gamma \ket{n+1-k}  } \\
& \geq \norm{ U(t)\ket{j} - \gamma \ket{n+1-j}  }
\end{aligned}
\end{equation}
for each $j=1,\ldots, m$. This proves that condition \eqref{eq:PGSTdef1} is satisfied. $\square$

We remark that, even though we have taken the magnetic field to be zero in the XX Hamiltonian \eqref{eq:Hfirst}, the results of Corollary 3 is also valid if a uniform magnetic field is added to the Hamiltonian. This is so because, in such case, the single-excitation transfer amplitudes would only be shifted by an overall phase factor \cite{analytical1}.

This result per se is already quite interesting, but the fact that we have restricted the analysis to the single-excitation subspace is a major limitation to its practical applicability. Hence, we will now turn our attention to the higher-excitation framework.

%=================================================================

\subsection*{Higher-excitation manifolds} The $r$-excitation subspaces ($r=2,3, \ldots, m$) are spanned by the set of states $ \ket{\ell_1, \ell_2, \ldots, \ell_r}$ (with $1 \leq \ell_s \leq n$ and pairwise distinct), where the qubits $\ell_1, \ell_2, \ldots, \ell_r$ are in the state $\ket{1}$ and the remaining qubits are in the state $\ket{0}$; the dimension of these subspaces is thus given by the binomial coefficients $\binom{n}{r}$. For example, $\ket{1,3,4}$ denotes the state $\ket{1_1 0_2 1_3 1_4 0_5 \ldots 0_n}$, which belongs to the $3$-excitation manifold.

In this unrestricted case, the sender wishes to transmit a state $\ket{\psi_{in}}$ which corresponds to a superposition of states with up to $m$ excitations. Therefore, the state of the whole system after placement of the input state is 
\begin{equation}
\begin{aligned}
\ket{\Psi (0)} = \: & \ket{\psi_{in}} \ket{0_{m+1} 0_{m+2} \ldots 0_{n}} \\
= \: & \alpha \ket{\underline{0}} + \sum_{j=1}^m \beta_j \ket{j} \\
& + \sum_{r=2}^m \left( \sum_{\ell_1, \ell_2, \ldots, \ell_r} \beta_{\ell_1 \ell_2 \ldots \ell_r} \ket{\ell_1, \ell_2, \ldots, \ell_r} \right)
\end{aligned}
\end{equation}
where the latter sum ranges over all the combinations of indices such that $1 \leq \ell_1 < \ell_2 < \ldots < \ell_r \leq m$, and where $\abs{ \alpha }^2 + \sum_{j=1}^{m} \abs{ \beta_j }^2 +  \sum_{r=2}^m \left( \sum_{\ell_1, \ell_2, \ldots, \ell_r} \abs{ \beta_{\ell_1 \ell_2 \ldots \ell_r} }^2 \right) = 1$. 

In this context, we must adapt our previous definition of PGST: we say that there is PGST of $\ket{\psi_{in}}$ if for any $\epsilon > 0$ there is a time $t>0$ such that
\begin{equation} \label{eq:PGSTdef2.1}
\abs{ \bra{n+1-j} U(t) \ket{j} } > 1-\epsilon, \qquad j=1,\ldots, m
\end{equation} and, for each $r=2, 3, \ldots, m$ and for each combination of indices $\ell_1, \ell_2, \ldots, \ell_r$,
\begin{equation} \label{eq:PGSTdef2.2}
\abs{ \bra{\widetilde{\ell_1}, \widetilde{\ell_2}, \ldots, \widetilde{\ell_r}} U(t) \ket{\ell_1, \ell_2, \ldots, \ell_r} } > 1-\epsilon,
\end{equation}
where $\widetilde{\ell_s}=n+1-\ell_s$. In other words, there is PGST of $\ket{\psi_{in}}$ if there is a solution $t>0$ for the nonlinear system of inequalities composed by the $m$ inequalities \eqref{eq:PGSTdef2.1} and by the $\binom{m}{2} + \ldots + \binom{m}{m}$ inequalities \eqref{eq:PGSTdef2.2}; this assures that all of the $2^m-1$ conditions are simultaneously verified at time $t$. This definition reduces to the former one in the single-excitation case, and it likewise implies that $\bra{\psi_{in}} \rho_{out}(t) \ket{\psi_{in}}$ can be made arbitrarily close to one.

Therefore, in this unrestricted framework we also have to consider conditions \eqref{eq:PGSTdef2.2}, which apparently represents a much harder challenge, especially because the transfer time must be the same for states with different number of excitations (i.e, the time $t$ cannot depend on $r$). However, we will show that the physical properties of the XX spin chain system imply that the evolution of higher excitation states can be easily determined through the analysis of the single-excitation manifold.

The key observation is that the XX spin chain is equivalent to a system of noninteracting fermions, since the Jordan-Wigner transformation allows us to write the Hamiltonian $H$ in the form \eqref{eq:Hferm}, where $c_j$ and $c_j^\dagger$ are fermionic operators. According to the Pauli exclusion principle, two noninteracting fermions cannot be simultaneously in the same quantum state at the same position. Hence, the $r$-fermion wave functions can be represented as an anti-symmetric product of the single-particle wave functions which can be formally expressed through the Slater determinant \cite{fermions, applied}.

The considerations above show that the composite qubits $\ket{\ell_1, \ell_2, \ldots, \ell_r}$ can be interpreted as states in which we have $r$ fermions occupying the positions $\ell_1, \ell_2, \ldots, \ell_r$. Accordingly, the Slater determinant formalism implies that the amplitudes of transfer for such fermionic states can be calculated as \cite{coherent, multiexcit}
\begin{equation} \label{eq:slater}
\begin{aligned}
\langle\widetilde{\ell_1}, \widetilde{\ell_2},& \ldots, \widetilde{\ell_r}| U(t) \ket{\ell_1, \ell_2, \ldots, \ell_r} = \\
= &\begin{vmatrix}
\bra{ \widetilde{\ell_1} }U(t)\ket{ \ell_1 } & \bra{ \widetilde{\ell_2} }U(t)\ket{ \ell_1 } & \cdots & \bra{ \widetilde{\ell_r} }U(t)\ket{ \ell_1 } \\
\bra{ \widetilde{\ell_1} }U(t)\ket{ \ell_2 } & \bra{ \widetilde{\ell_2} }U(t)\ket{ \ell_2 } & \cdots & \bra{ \widetilde{\ell_r} }U(t)\ket{ \ell_2 } \\
\vdots & \vdots & & \vdots \\
\bra{ \widetilde{\ell_1} }U(t)\ket{ \ell_r } & \bra{ \widetilde{\ell_2} }U(t)\ket{ \ell_r } & \cdots & \bra{ \widetilde{\ell_r} }U(t)\ket{ \ell_r } \\
\end{vmatrix}.
\end{aligned}
\end{equation}
We use as a lemma the following result of Ostrowski \cite{determinant}:

\emph{Lemma 4.} Let $A=[a_{ij}] \in \Cc^{r\times r}$ be a complex matrix such that $\abs{a_{ii}} > \sum_{j\neq i} \abs{a_{ij}}$. Then,
\begin{equation}
\abs{\det{A}} \geq \prod_{i=1}^r \left( \abs{a_{ii}} - \sum_{j\neq i} \abs{a_{ij}} \right).
\end{equation}

In order to explain our reasoning, let us now consider one of the inequalities \eqref{eq:PGSTdef2.2}, i.e, we regard $\ell_1, \ell_2, \ldots, \ell_r$ as fixed. Denote the elements of the matrix in the right hand side of \eqref{eq:slater} by $a_{ij}$. (Note that the $a_{ij}$ depend on $t$.) By Corollary 3, for any given small $\delta > 0$, there is a time $t_\delta$ such that $\abs{a_{ii}} > 1-\delta$ and \begin{equation}
\begin{aligned}
\sum_{j \neq i} \abs{a_{ij}} & \leq (r-1) \max_{j\neq i} \{ \abs{a_{ij}} \} \\
& \leq (r-1)\sqrt{2\delta - \delta^2} \\
& < (r-1)\sqrt{2\delta},
\end{aligned}
\end{equation}
for all $i$. \footnote{Recall that, since $U(t_\delta)$ is unitary, each of its rows satisfies the normalization condition, whence $\max_{j\neq i} \{ \abs{a_{ij}}\} \leq \sqrt{2\delta - \delta^2}$. Observe also that $\delta>0$ is small and thus $0 < 2\delta - \delta^2 < 2\delta$.} By Lemma 4, we obtain 
\begin{equation}
\begin{aligned}
\Big| \bra{\widetilde{\ell_1}, \widetilde{\ell_2},  \ldots, \widetilde{\ell_r}} U(t_\delta) &\ket{\ell_1, \ell_2, \ldots, \ell_r} \Big| > \\
& > \left( 1 - \delta - (r-1)\sqrt{2\delta} \right)^r \\
& > \left( 1 - r\sqrt{2\delta} \right)^r. \\
\end{aligned}
\end{equation}
If we choose $\delta = \left( (1-\epsilon)^{1/r} - 1 \right)^2  / \left(2r^2\right) >0 $ , we obtain \eqref{eq:PGSTdef2.2}. Though the respective value of $t_\delta$ depends on which of the inequalities \eqref{eq:PGSTdef2.2} we are considering, Corollary 3 assures that there is a time $t_\delta^* > 0$ at which the smallest of such values of $\delta$ is achieved, and it is easily seen that every one of the inequalities \eqref{eq:PGSTdef2.2} is simultaneously verified at time $t_\delta^*$. Thus, we have proved the following theorem, our main result:

\emph{Theorem 5.} Suppose the Hamiltonian of the chain is given by \eqref{eq:Hfirst}. Let $\ket{\psi_{in}}$ be an arbitrary composite $m$-qubit state. Then, there is PGST of the state $\ket{\psi_{in}}$ if and only if $n=p-1, 2p-1$ or $2^k-1$, where $p$ is a prime and $k \in \Nn$.

Note that, since $\abs{ \bra{\widetilde{\ell_1}, \widetilde{\ell_2}, \ldots, \widetilde{\ell_r}} e^{-iHt} \ket{\ell_1, \ell_2, \ldots, \ell_r} }$ are the amplitudes for the mirror inversion of states in the $r$-excitation subspace, our arguments show that the pretty good mirroring of single-particle states is a sufficient condition for the pretty good mirror inversion of the state of the register in each excitation subspace. Consequently, the problem of the PGST of multi-particle states can be reduced to the analysis in the single-qubit transfer scenario, and this is true not only for the case of the unmodulated XX Hamiltonian but also for any excitation-preserving Hamiltonian. These results are consistent with those reported in \cite{mirrorreg, spinmirror} for the case of PST.

We now consider the practical scenario where the tolerance $\epsilon$ is fixed \textit{a priori} to a specific small value. Fig.\ 2 shows the first times for which there is high fidelity, for various chain lengths $n$ and tolerances $\epsilon$. In Fig.\ 2(a), where we only consider chain lengths for which there is PGST, the data indicates that the transfer times grow exponentially as the length of the chain increases. The second plot shows a stairlike behavior of the transfer times as the tolerance $\epsilon$ decreases, which is due to the fact that, as the system evolves in time, the fidelity shows various peaks of different intensities. In both cases, the numerical evidence suggests that the transfer times show only a modest increase as the size of the multi-qubit state we wish to transfer gets larger. Our scheme thus offers the possibility to transfer multi-particle states at short distance, e.g. between adjacent quantum processors \cite{bose1}.

\begin{figure}[h] \label{figure2}
     \centering
     \makebox[8cm][c]{
     \renewcommand{\tabcolsep}{1pt}
     	\begin{tabular}{c}    	
     	\vspace*{-2pt}  \includegraphics[width=8cm]{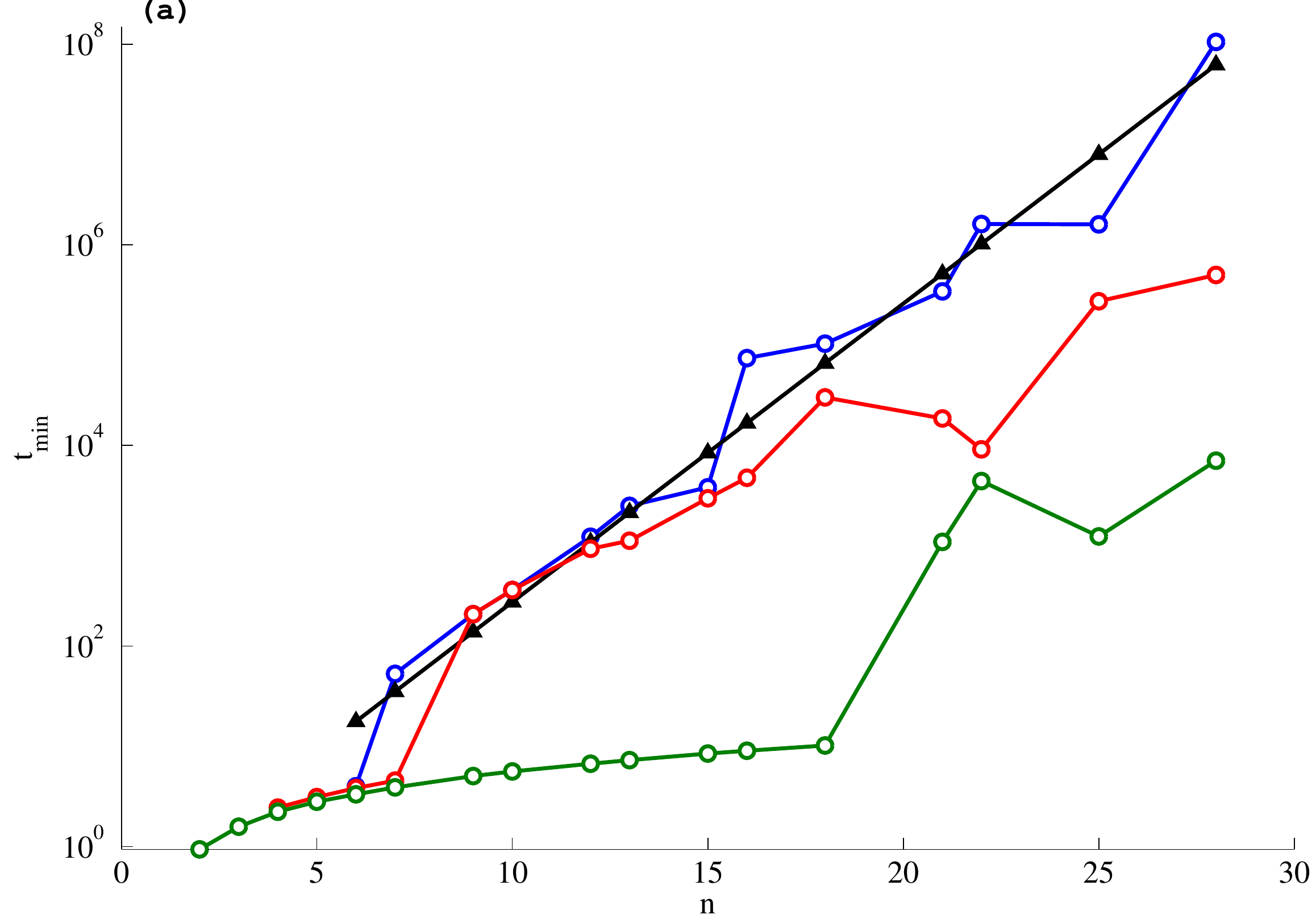}\\
     	\vspace*{-2pt}   \includegraphics[width=8cm]{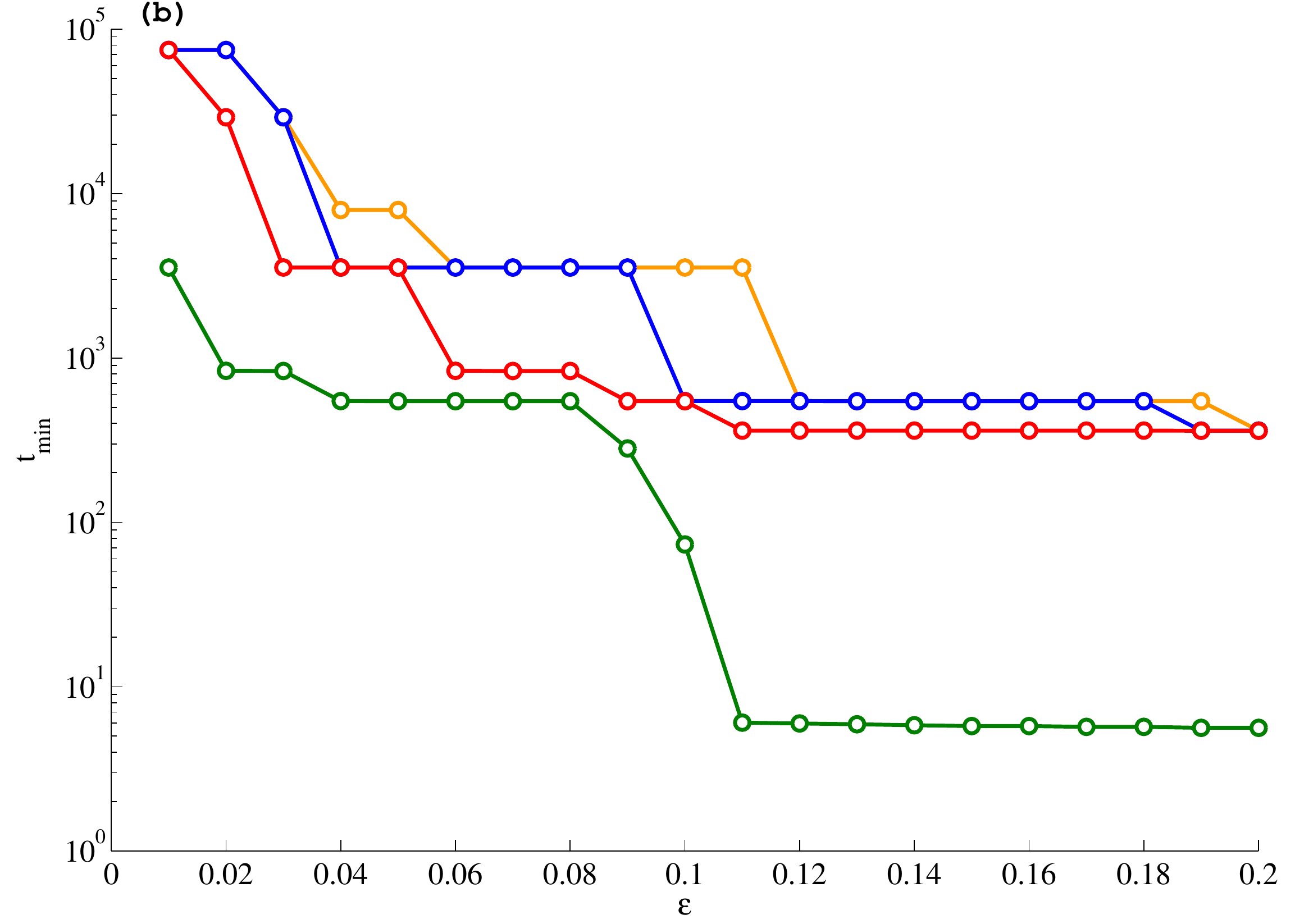}\\
        \end{tabular}
     }
     \caption{Semi-logarithmic plot of $t_{min}$, which is the smallest value of $t$ satisfying equations \eqref{eq:PGSTdef2.1} and \eqref{eq:PGSTdef2.2}, (a) as a function of the chain length $n$, for a fixed tolerance $\epsilon=0.2$; and (b) as a function of $\epsilon$, for a fixed chain length $n=10$. In both panels, the green, red and blue lines correspond to the cases of states of $m=1$, $m=2$ and $m=3$ qubits respectively; in panel (b) we have also included the case of states of $m=4$ qubits (orange line). In panel (a), the black line represents the best least-squares exponential fit to the data for $m=3$ qubits, which is given by $t_{min} \simeq 0.29 e^{0.6852n}$.}
\end{figure}

Ultracold atoms in optical lattices are a promising candidate for the experimental verification of our results. In ensembles of cold atoms, where the implementability of the unmodulated XX Hamiltonian has been demonstrated \cite{opticalham}, the ordinary values of the coupling constant $J$ are hundreds of Hertz (for instance, the value of $J = 360h$ Hz was reported in \cite{opticalcoupl}). For a tolerance $\epsilon = 0.1$, the minimal time for transferring a state of $m=3$ qubits in a chain with $n=10$ sites is $Jt_{min}/\hbar \simeq 546$ (see Fig.\ 2(b)), which yields $t_{min} \simeq 0.24$ s. This is manifestly compatible with current technology, since evolution up to 0.21 s has been reported in \cite{optical1, optical2}, while other experiments have observed coherence times of $\simeq 9$ min \cite{opticaltime}. Though the exponential transfer times hinder the scalability of the protocol, this indicates that unmodulated chains represent a simple practical solution for short-distance quantum state transmission, useful for the design of quantum information devices.

%=================================================================

\section{Conclusions}

We have proved that any multi-qubit state can be transferred with arbitrarily large fidelity through the uniform XX quantum spin chain if and only if the length of the chain is $n=p-1$, $n=2p-1$, or $n=2^k-1$. This is a significant generalization of the results of \cite{pgst} for single-particle transfer, and shows that the simplest spin chain model for quantum state transfer can be used as a data bus for quantum information processing devices. We have also demonstrated that, for any excitation-preserving Hamiltonian, the study of the PGST of multi-qubit states can be reduced to the corresponding single-qubit problems. This significantly simplifies future research on PGST.

While the application of quantum algorithms to problems in number theory (and, in particular, primality) is a long-established and prominent investigation topic \cite{algorithms}, the connection between non-algorithmic quantum dynamics and prime numbers has not yet drawn much attention from researchers. By expanding the scope of scenarios in which the fidelity of state transfer is characterized as a function of primality constraints, our work corroborates the viewpoint that quantum spin systems can potentially be used as a tool for tackling problems in prime number theory \cite{pgst}.

Finally, our results indicate that the choice of a uniform coupling configuration can represent a simple practical solution for the design and implementation of universal short-distance quantum data buses, capable of transferring arbitrary quantum states with very high fidelity. Furthermore, we showed how the current capability to manipulate ultracold atoms in optical lattices allows for the experimental implementation of our protocol. Other potential candidates for the experimental verification of our protocol are nuclear magnetic resonance \cite{expuniform, nmrliquid}, photonic waveguide lattices \cite{photonic1, photonic2} and atomic nanomagnets \cite{nanomagnets}, given the substantial recent accomplishments in the experimental realization of quantum spin chains in these platforms.

%=================================================================

\begin{acknowledgments} The authors are grateful for the support from Fundação para a Ciência e a Tecnologia (Portugal), namely through programmes PTDC/POPH and projects PEst-OE/EGE/UI0491/2013, PEst-OE/EEI/LA0008/2013, IT/QuSim, and CRUP-CPU/CQVibes, partially funded from EU FEDER, and by the EU FP7 projects LANDAUER (GA 318287) and PAPETS (GA 323901).
\end{acknowledgments} 

%====================================================

%\bibliographystyle{apsrev4P}

\end{document}